\documentclass[twocolumn,showpacs,preprintnumbers,amsmath,amssymb]{revtex4-1}

\usepackage{graphicx}% Include figure files
\usepackage{dcolumn}% Align table columns on decimal point
\usepackage{bm}% bold math
\usepackage{color}
\usepackage{soul}

\begin{document}

\title{Quantum confinement of zero-dimensional hybrid organic-inorganic polaritons at room temperature }

\author{H. S. Nguyen$^1$}
\author{Z. Han$^{1,2}$}
\author{K. Abdel-Baki$^2$}
\author{X. Lafosse$^1$}
\author{A. Amo$^1$}
\author{J-S. Lauret$^2$}
\author{E. Deleporte$^2$}
\author{S. Bouchoule$^1$}
\author{J. Bloch$^1$}
\email{jacqueline.bloch@lpn.cnrs.fr}
\affiliation{$^1$Laboratoire de Photonique et de Nanostructures, LPN/CNRS, Route de Nozay, 91460 Marcoussis , France}
\affiliation{$^2$Laboratoire Aim\'e Cotton, \'Ecole Normale Sup\'erieure de Cachan, CNRS, Universit\'e Paris Sud, bat.505 campus d'Orsay, 91405 Orsay, France}

\date{\today}

\begin{abstract}
We report on the quantum confinement of zero-dimensional polaritons in perovskite-based microcavity at room temperature. Photoluminescence of discrete polaritonic states are observed for polariton localized in symmetric sphere-like defects which are spontaneously nucleated on the top dielectric Bragg mirror. The linewidth of these confined states are found much sharper (almost one order of magnitude) than that of photonic modes in the perovskite planar microcavity. Our results show the possibility to study organic-inorganic cavity polariton in confined microstructure and suggest a fabrication method to realize integrated polaritonic devices operating at room temperature.
\end{abstract}

\maketitle
Over the last decade, cavity polaritons - quasiparticles arising from the strong coupling regime between quantum well excitons and microcavity photons \cite{Weisbuch}, have emerged as an important research topic for both fundamental and applied physics. On one hand, cavity polaritons provide an interesting insight for studying the physics of Bose-Einstein condensate in an out of equilibrium system \cite{Deng,Kasprazak,Bajoni,Amofluid,Lagoudakis}. On the other hand, polariton self-interaction \cite{Baas} and ballistic propagation of polariton condensate in engineered confined-microstructures \cite{Freixanet,Wertz} reveal a new generation of optical devices \cite{Liew,Shelykh:2009,Amoswitch,Gao,Nguyen,Ballarini,Sturm}. State-of-the-art microcavities are based on GaAs compounds where Q-factors as high as $100000$ can be obtained \cite{Nguyen,Langbein}, with the facilities of engineering the polariton potential by different techniques \cite{Bloch,Kaitouni,Lai,Cerda,Wertz,AmoPRB,Tosi}. Nevertheless, due to the small exciton binding energy in GaAs, their operating regime is limited to cryogenic temperatures.  The search for room temperature polaritonic system is therefore oriented into materials having larger exciton binding energy such as GaN \cite{Christopoulos,Christmann}, ZnO \cite{Li}, organic and hybrid organic/inorganic materials \cite{Lidzey1,Lidzey2,KenaCohen,Plumhof,Lanty}. As a matter of fact, although Bose-Eistein condensation of cavity polaritons at room temperature have been recently observed \cite{Christopoulos,Christmann,Li,KenaCohen,Plumhof}, no polaritonic microstructures at room temperature has been reported so far due to  difficulties associated with the patterning process of GaN and ZnO, as well as the fragility of organic materials. This limit is therefore an obstacle for the prospective of integrated polaritonic device operating at room temperature.

\begin{figure}[htb]
\begin{center}
\includegraphics[width=8cm]{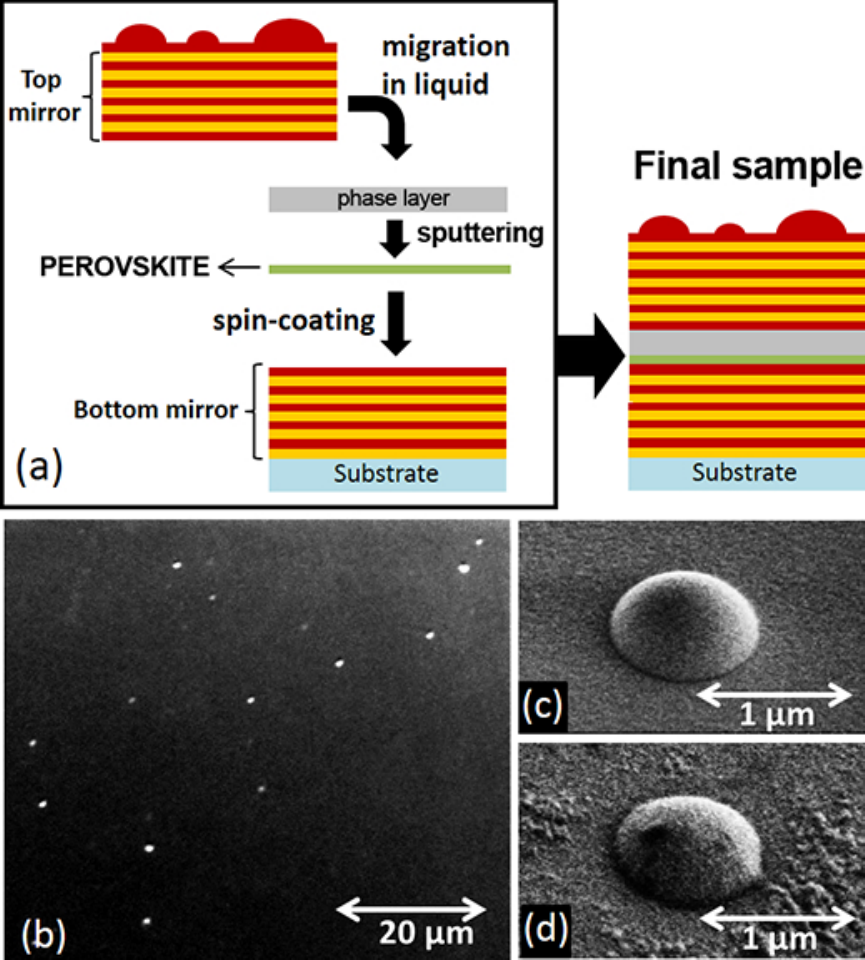}
\end{center}
\caption{\label{fig1}{(a) Different steps of the sample fabrication. (b) Optical microscopy image of an ensemble of sphere-like defects on the top Bragg mirror. (c) SEM image of a sphere-like defect in the top Bragg mirror before the release from its substrate, having a diameter of 900 nm and height of 300 nm. (d) SEM image of a sphere-like defect of the top Bragg mirror on the final sample.}}
\end{figure}

\begin{figure*}[htb]
\begin{center}
\includegraphics[width=14cm]{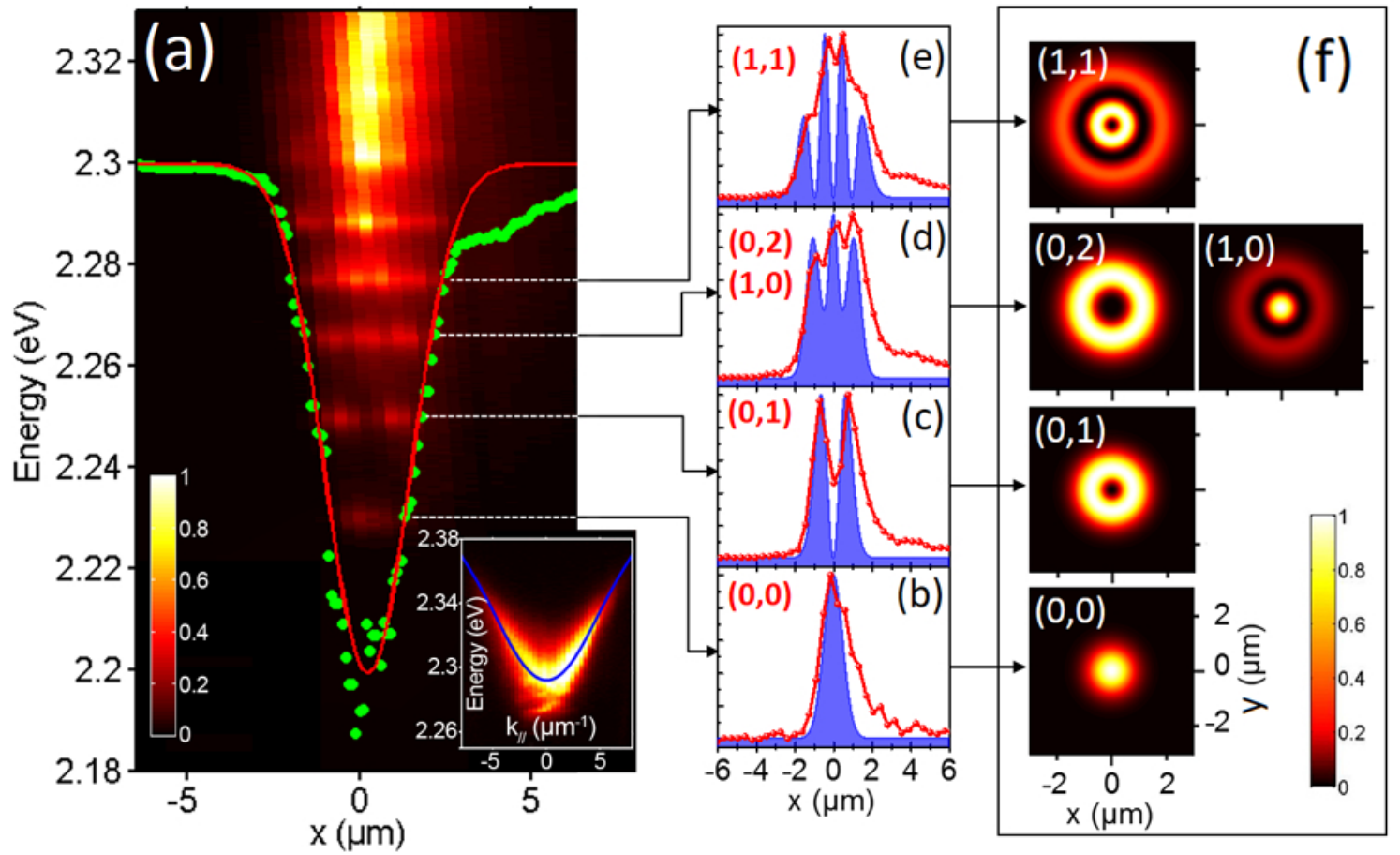}
\end{center}
\caption{\label{fig2}{(a) Photoluminescence intensity $I(x,E)$ of polariton emission within defect D1 as a function of position and energy. Green circles: polariton potential of confinement within D1. Red line: Gaussian fit of FWHM amounts to $2.6\,\mu m$. Inset: Far field emission of 2D polaritons measured in the planar region outside D1 of detunning $\delta=-36\,meV$. (b)-(e) Measured (connected red circles) and simulated (blue shade) photoluminescence intensity of different confined modes as a function of the position along x-axis.(f) Simulation of the spatial intensity distribution of the first five eigenstates calculated with the fitting Gaussian potential shown in (a).}}
\end{figure*}

In this letter, we demonstrate the quantum confinement of zero-dimensional (0D) perovskite-based polaritons in sphere-like defects, which are spontaneously formed during the fabrication of the top dielectric mirror. This top mirror is fabricated separately from the rest of the sample and then attached to the spin-coated perovskite layer by migration technique in liquid \cite{Zheng_APEX}. Microphotoluminescence measurements performed on single defects at room temperature show up to 7 states of discrete polaritons in a $100\,meV$ confinement potential. Numerical resolution of a Schr\"odinger equation provides a good agreement with experimental measurements for the spatial distribution and the energy confinement of these states. Interestingly, the linewidth of the confined states is much sharper than that of photonic modes in the perovskite planar microcavity \cite{Zheng}. We attribute this effect either to a local increase of the cavity thickness due to the defect nucleation, or to a suppression of inhomogeneous broadening of the cavity mode within the micron-size defect.

The fabrication method of our sample is similar to the one reported in \cite{Zheng_APEX}, and is schematically presented in Fig.~\ref{fig1}(a). First, the bottom dielectric Bragg mirror (11 pairs of $SiN_x/SiO_2$ layers) is deposited on a fused silica substrate by Plasma-Enhanced Chemical Vapor Deposition (PECVD). Second, a 60 nm-perovskite layer (hybrid organic-inorganic material of chemical formula $(C_6H_6C_2H_4-NH_3)_2PbI_4$ \cite{Lanty}), is spin-coated on the bottom Bragg mirror and forms a multi-quantum well structure. This layer is then covered by a $SiN_x$ layer to complete the $\lambda/2$ cavity. The top dielectric Bragg mirror (8 pairs of $YF_3/ZnS$) is deposited by vacuum evaporation at low temperature ($<\,100^oC$) onto a sacrificial layer. After releasing the top mirror by dissolving the sacrificial layer, we position it onto the perovskite layer using a migration technique in liquid (toluene). Finally, the whole structure is annealed  to eliminate the residual liquid and to firmly attach the two surfaces. The Q-factor of this $\lambda/2$ microcavity measured by reflectivity and photoluminescence experiments amounts to 80, with a Rabi splitting at room temperature of $\Omega=168\,meV$ as reported in the reference \cite{Zheng}. The sample presents a modulation of the energy detuning $\delta = E_C -E_X$ between the photonic mode and the perovskite exciton resonance due to a thickness modulation of the spin-coated perovskite layer. $\delta$ varies typically from $-150\,meV$ to $50\,meV$ over a distance of $120\,\mu m$.  

%The sample presents a gradient of the energy detuning $\delta = E_C -E_X$ between the photonic mode and the perovskite exciton resonance. $\delta$ varies typically from $-150\,meV$ to $50\,meV$ over a distance of $120\,\mu m$. This gradient is due to a thickness modulation of the spin-coated perovskite layer.

During the evaporation process performed at low temperature for the fabrication of the top Bragg mirror, sphere-like defects are spontaneously formed. A cross-sectional inspection of a single sphere is required to determine the physical origin of the defect formation. Such analysis relies on specific focused ion beam technology and would destroy the sample.  We believe that a weak surface adhesion combined to the strain accumulated into the dielectric films may induce a local detachment of the film from the surface. It is also likely that the formation of spheres is assisted by a degassing process occurring during the evaporation of the dielectric materials (similar to the formation of hydrogen micro-cavities in amorphous hydrogenated silicon deposited by PECVD using $SiH_4$ \cite{Lebib}). These defects have a symmetric half-sphere shape with diameter varying from a few hundred nanometers to a few micrometers measured by Scanning Electronic Microscopy (SEM). Their density is very low, around $10^{-2}\,\mu m^{-2}$ but we always observe ensemble of defects separated by less than $20\,\mu m$ (see Fig.~\ref{fig1}(b)). A SEM image of a typical defect on the surface of the initial top mirror before the release from its substrate is shown in fig.~\ref{fig1}(c). Once the top mirror is released and attached to the perovskite layer, the sphere-like defects are fully conserved on the final samples as shown in Fig.~\ref{fig1}(d). Notice that similar defects have been reported for GaAs based cavities \cite{Langbein}. In this case, they were attributed to point-like defects formed during the molecular beam epitaxy of the top mirror.

Confocal microphotoluminescence on single sphere-like defects is performed at room temperature, using a laser beam (405 nm) focused onto a $1.5\,\mu m$ spot diameter  by a microscope objective ($NA=0.65$). Polariton emission is imaged on a CCD camera coupled to a monochromator.

We first report the study of polariton emission of a representative sphere-like defect in our sample, denoted D1. Figure.~\ref{fig2}(a) presents the photoluminescence intensity $I(x,E)$ of polariton emission within D1 as a function of position and energy. We distinguish five discrete polariton modes, which correspond to 0D polaritons within the defect due to confinement of the photonic component. We use the method  proposed by Zajac \textit{et al} \cite{Langbein} to calculate the potential confinement of these polaritons. Our hypothesis is that D1 is perfectly symmetric and all physical quantities only depend  on the radial distance $r$. We denote by $\Psi_n(r)$, the wave function of the $n^{th}$ confined state and by $I_n(r)$ its photoluminescence emission intensity. $I_n(r)$ is extracted directly from the intensity of polariton emission in real space $I(x,E)$: $I_n(r) \propto I(\left|x\right|=r,E=E_n)$. The spectrally integrated density of confined states $D(r)$ is given by:
\begin{equation}
D(r) = \sum\limits_{n=1}^5 \left|\Psi_n(r)\right|^2 = \sum\limits_{n=1}^5 \frac{I_n(r)}{2\pi\int_{0}^{6\,\mu m} I_n(r')r'\,dr'} \label{Dm}
\end{equation}
Once $D(r)$ is obtained, the potential of confinement $V(r)$ is simply calculated using \cite{Langbein}:
\begin{equation}
V(r) = \frac{2\pi\hbar^2}{m_{2D}}\times D(r)\label{V}
\end{equation}
where $m_{2D}$ is the mass of free polaritons outside D1 ($m_{2D}=1.5\times10^{-5}m_{electron}$), obtained by fitting the dispersion of the flat region in the neighborhood of D1 (see inset of fig.~\ref{fig2}(a)). The calculated $V(r)$ is well fitted by a Gaussian dip of amplitude $103\,meV$ and of full width at half maximum (FWHM)  $2.6\,\mu m$ [see fig.~\ref{fig2}(a)].

\begin{figure}[htb]
\begin{center}
\includegraphics[width=6cm]{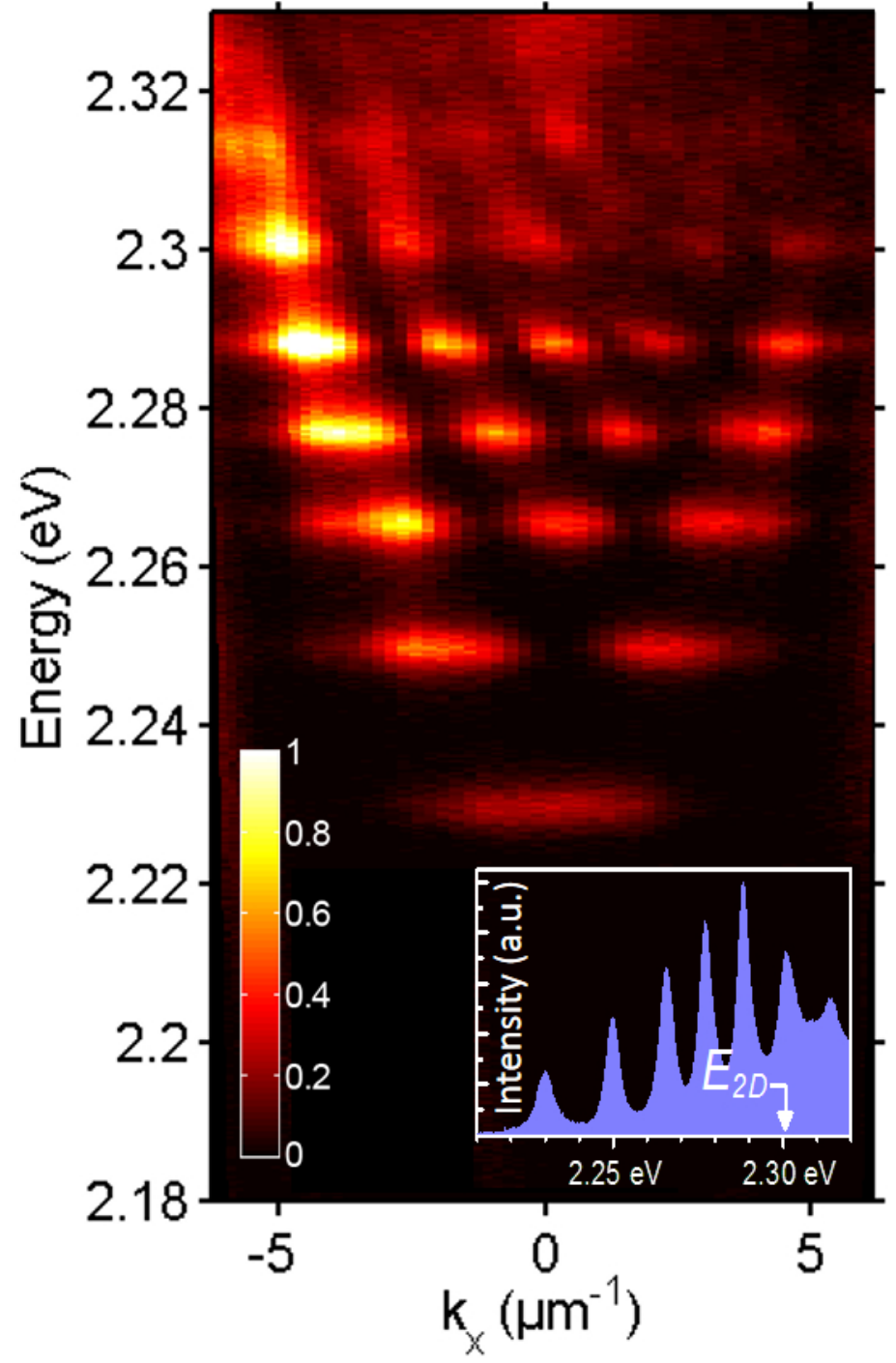}
\end{center}
\caption{\label{fig3}{Far field emission intensity measured in D1. Inset: Integrated photoluminescence intensity as a function of energy.}}
\end{figure}

We then solve numerically the 2D Schr\"{o}dinger equation for a single particle in presence of this potential. Each eigenstate of such potential is defined by two quantum numbers $(n_r,l)$, where $n_r$ is the number of nodes of the radial wave function and $l$ is the angular momentum number. Our simulation shows that the first-five eigenstates are $(0,0),(0,1),(1,0),(0,2),(1,1)$ where $(1,0)$ and $(0,2)$ are degenerate. The spatial emission patterns of these states are presented in fig~\ref{fig2}(f). As shown in figs~\ref{fig2}(b)-(e), our numerical simulations of the spatial distribution of the first four discrete modes are in good agreement with the experimental results, which confirm our hypothesis of the symmetry of D1. However, due to a more complex structure of the real confinement potential (presence of a tunneling barrier with respect to the 2D region as discussed at the end of this paper), our simulation cannot reproduce the fifth discrete mode which is close to the 2D region. %Nevertheless, by choosing $m=0.94\times10^{-5}m_{electron}$, we can even reproduce successfully the energy of the first four discrete modes. Our estimated confinement potential $V(r)$ is therefore reliable to simulate the confined modes well below the 2D continuum.

\begin{figure}[htb]
\begin{center}
\includegraphics[width=8cm]{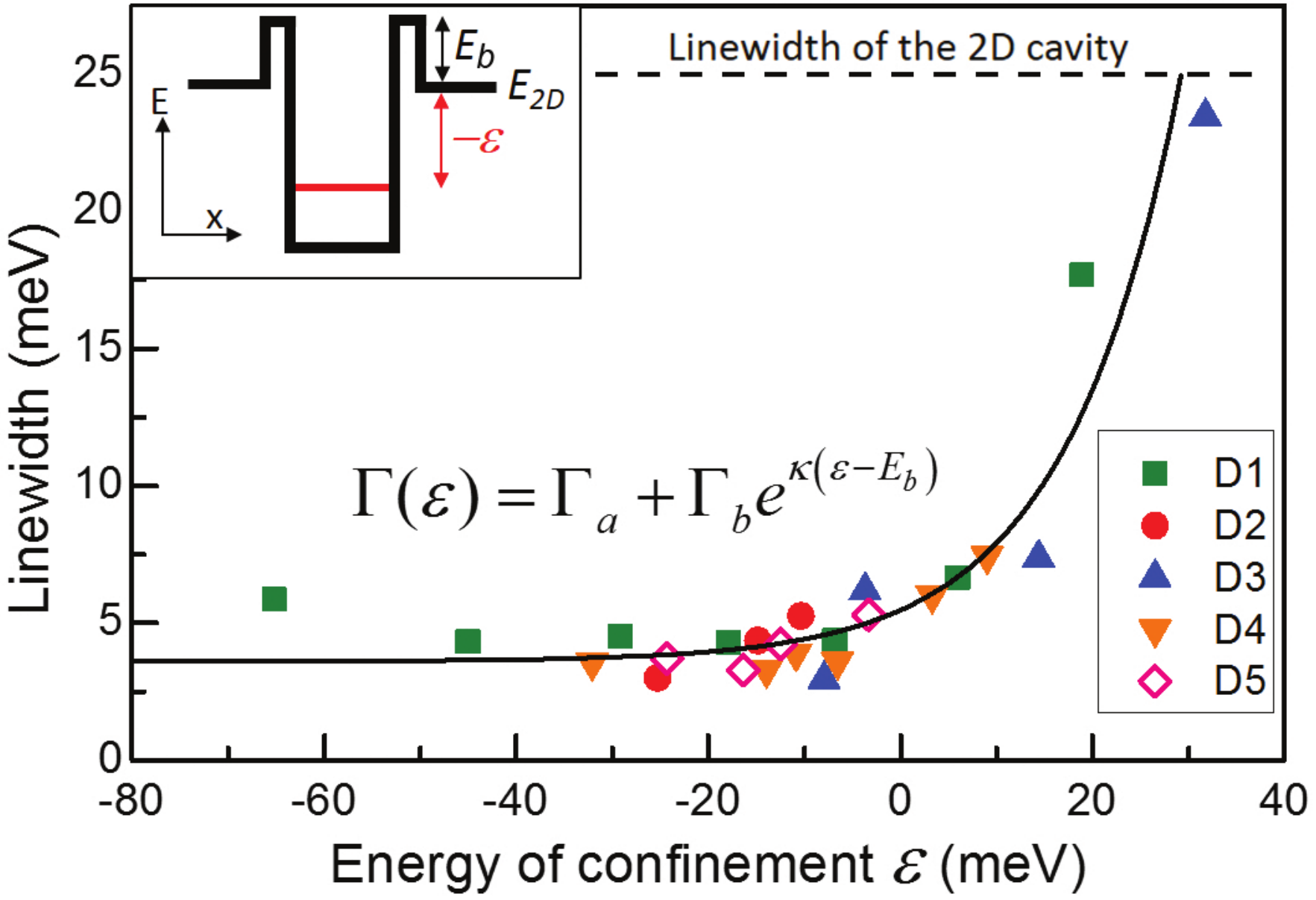}
\end{center}
\caption{\label{fig4}{Linewidth of discrete states measured on five defects (D1$\rightarrow$D5) as a function of the energy of confinement $\epsilon = E_{2D} - E_n$. Inset: Sketch of the confinement potential with the presence of a tunneling barrier between the defect region and the 2D continuum.}}
\end{figure}

The far-field image of the polariton emission within D1 is presented in fig.~\ref{fig3}. We distinguish in this figure seven discrete polariton modes -  two more than the number observed in the real space spectrum [i.e. fig.~\ref{fig2}(a)]. In fact, in real-space emission, the last two discrete modes are covered by a strong continuum of emission  attributed to a third polariton branch resulting from the strong coupling between the perovskite exciton and a Bragg mode \cite{Zheng}. This emission occurs at very large wave-vector and is thus filtered from the far-field  image. This is why more modes are imaged in fig.~\ref{fig3}. Interestingly, the sixth and seventh modes observed in fig.~\ref{fig3} are above the bidimensional continuum ($E_{2D}=2.3\,eV$), suggesting the presence of a potential barrier between the defect region and the neighborhood 2D area \cite{Langbein} (see the Inset of fig.~\ref{fig4}).

Finally, we would like to discuss the effect of quantum confinement on polariton linewidth. Figure.~\ref{fig4} presents the measured linewidth of discrete states measured in five different defects, as a function of the energy of confinement $\epsilon = E_{2D} - E_n$. These results show that all the confined states (i.e. $\epsilon<0$) have a linewidth 4-8 times smaller than the one of the 2D photonic mode ($\Gamma_{2D\,cavity} = 25\,meV$). In particular, the first confined state of D2 has a linewidth of $3\,meV$, corresponding to a quality factor $Q\geq750$ (Q=750 if this state is $100\%$ photon-like). This value is one order of magnitude higher than the Q-factor of the perovskite-based planar cavity measured on the same sample \cite{Zheng}. Two physical reasons can be involved to explain these narrow lines. First, the nucleated defects could increase locally the thickness of the cavity, thus increasing locally the Q-factor since the quality factor is proportional to the length of the cavity.  Second, similar enhancement effect has been observed previously in GaAs based microcavity mesa structures at cryogenic temperature\cite{Kaitouni}, and was explained by the suppression of inhomogeneous broadening due to long-range cavity thickness fluctuations. Cross-sectional SEM or Scanning Transmission Electron Microscopy investigation would be required to discriminate between these two interpretations. Nevertheless, the polariton modes, well protected from the excitonic reservoir, do not appear to be broadened by phonon scattering \cite{Trichet}. This is why we can obtained very narrow polariton modes despite room temperature operation. The dependance of the linewidth as a function of the confinement energy is well fitted by (solid curve in fig.~\ref{fig4}):
\begin{equation}
\Gamma(\varepsilon)=\Gamma_a + \Gamma_b e^{\kappa(\varepsilon-E_b)}\label{Gamma}
\end{equation}
with $\Gamma_a=3.6\,meV$, $\Gamma_b=21\,meV$, $E_b=29\,meV$ and $\kappa=0.08\,meV^{-1}$. We can interpret the second term of eq.(3) as the tunneling rate of confined polaritons to the 2D region through a barrier potential $E_b$ which isolate the defect region with the surrounding area (see the Inset of fig.~\ref{fig4}).

In conclusion, zero-dimensional cavity polaritons have been demonstrated at room temperature, making use of polariton confinement within micron-size sphere-like defects. Interestingly, the achievement of 0D polaritons can be transposed to other types of active media as the formation of the sphere-like defect is independent from the deposition of the active medium. Moreover, more complicated microstructures could be engineered at will, using lithography \cite{Wertz,Zhang,Ding}and patterning techniques applied to the top mirror instead of spontaneous nucleated defects.  Our work thus opens a way to realize integrated polaritonic devices operating at room temperature for all most any material.

This work is supported by Agence Nationale de la Recherche (ANR) within the Project PEROCAI, by R\'eseau Th\'ematique de Recherche Avanc\'ee (RTRA) Triangle de la Physique within the Project CAVPER and by the french RENATECH network.

\end{document}